\documentclass[a4paper,11pt]{article}

\usepackage[dvips]{graphicx}
\usepackage{amsfonts, color, float, amsmath}
\usepackage{amssymb}

\usepackage{url}
\usepackage{array}
\usepackage{textcomp}

\usepackage{slashed}
\usepackage[dvipsnames]{xcolor}

\usepackage[small,bf,sf]{caption}
\usepackage{microtype}
\usepackage{cite}
\usepackage{authblk}

\usepackage[bottom]{footmisc}
\usepackage{booktabs}
\usepackage{hyperref}

\usepackage{subfig}

\usepackage{colortbl}
\definecolor{myblue}{rgb}{0.8,0.85,1}
\definecolor{light-gray}{gray}{0.95}

\hypersetup{colorlinks=true,
citecolor=ForestGreen,
anchorcolor=CornflowerBlue,
linkcolor=NavyBlue,
urlcolor=Mahogany,
linkbordercolor={1 1 0}}

\allowdisplaybreaks

\def\beq {\begin{equation}}
\def\eeq {\end{equation}}
\def\bea {\begin{eqnarray}}
\def\eea {\end{eqnarray}}

\AtBeginDocument{}

\usepackage{listings}
\usepackage{color}

\definecolor{dkgreen}{rgb}{0,0.6,0}
\definecolor{gray}{rgb}{0.5,0.5,0.5}
\definecolor{mauve}{rgb}{0.58,0,0.82}

\usepackage{lipsum}

\title{\Large {\bf \sffamily \boldmath $H\to \gamma\gamma$ to all orders in $\alpha_s$ in the large-$\beta_0$ limit of QCD}}

\author[a]{D. Boito}
\author[a]{G. das Neves}
\author[b]{J. Piclum\vspace{0.3cm}}

\affil[a]{\it Instituto de F\'isica de S\~ao Carlos, Universidade de S\~ao Paulo, CP 369, 13560-970, S\~ao Carlos, SP, Brazil}

\affil[b]{\it Theoretische Physik 1,Center for Particle Physics Siegen (CPPS),\newline Universit\"at Siegen, Walter-Flex-Str. 3, 57068 Siegen, Germany}

\begin{document}
{\small \textsc{SI-HEP-2022-25, P3H-22-093} \hfill \today}

\vspace*{-0.2cm}
\begingroup
\let\newpage\relax
\maketitle
\endgroup
\date{}

\vspace*{-1.0cm}
\begin{abstract}

\noindent

\noindent In this note, we present the result for the QCD corrections to
the decay of the Higgs boson into two photons in the large-$\beta_0$ limit of QCD, providing the first two terms in the heavy-top expansion.
From our results, one can easily read off the exact leading-$n_f$ QCD contributions in analytic form
to all orders in the strong coupling, $\alpha_s$, where $n_f$ is the number of massless quarks, and identify the leading renormalon singularities. We give explicit results for the leading-$n_f$ coefficients at 6 and 7 loops and use the large-$\beta_0$ result to
speculate about the size of yet unknown (but small) higher-order contributions to the QCD series.
\end{abstract}

\thispagestyle{empty}

\setcounter{page}{1}

\section{Introduction}

The Higgs decay into two photons played a central
role, ten years ago, in the discovery of the Higgs
boson at the Large Hadron Collider (LHC). This channel remains crucial for precision studies of Higgs
properties at the LHC as well as at future
colliders. Given the expected increase in precision
on the experimental side, the theoretical calculation
of Higgs decays within the Standard Model (SM) must be performed at higher orders in perturbation theory.

Because the photon is massless, this
decay is a loop-induced process at leading order (LO) within the SM. Its
amplitude can be decomposed into a bosonic
contribution, stemming from the $W$ boson, and
fermionic contributions. Here we are concerned with
the fermionic contribution arising from the top quark, which dominates over the subleading $b$-quark
and $\tau$-lepton fermion loops (contributions from
even lighter fermions are irrelevant).

The decay amplitude has been known at LO for many years \cite{Ellis:1975ap, Shifman:1979eb, Okun:1982ap, Gavela:1981ri}. The next-to-leading order (NLO) result was first obtained as an expansion in $\tau_t
=\frac{m_H^2}{4m_t^2}\approx 0.13$, with $m_H$ and $m_t$ being the Higgs and the top-quark masses, respectively, but later the full $m_t$ dependence of the NLO result was obtained\cite{Djouadi:1990aj,Melnikov:1993tj,Djouadi:1993ji,Inoue:1994jq,Spira:1995rr,Fleischer:2004vb,Harlander:2005rq,Aglietti:2006tp}.
The NNLO QCD contribution, which is a three-loop calculation, was obtained about ten years ago, again as an asymptotic expansion in the top mass; numerical results with full top-quark mass dependence were obtained only much more recently\cite{Steinhauser:1996wy, Maierhofer:2012vv,Niggetiedt:2020sbf}. The NNNLO (four-loop), $\mathcal{O}(\alpha_s^3)$, corrections were first partially computed in Ref.~\cite{Sturm:2014nva} and later completed in Ref.~\cite{Davies:2021zbx}. Partial results for the N$^4$LO contributions where both photons couple to a massive top-quark loop can also be found in Ref.~\cite{Sturm:2014nva}.

In this work we present, for the first time, the result for the QCD corrections in the large-$\beta_0$ limit of QCD within the heavy-top limit.
From our results, the exact leading-$n_f$ QCD contributions, i.e. coefficients of terms proportional to $\alpha_s^n n_f^{n-1}$, with $n_f$ being the number of active (massless) quark flavors, can be directly read off in analytic form to all orders in the strong coupling.
In our calculation, we are concerned with non-singlet diagrams only, i.e. diagrams in which the Higgs boson and the two photons in the final state couple to the same top-quark loop. Contributions from the singlet diagrams are subleading in the large-$\beta_0$ power counting, and do not contribute to the terms $\mathcal{O}(\alpha_s^n n_f^{n-1})$.\footnote{These contributions are not small in full QCD and we comment on singlet-diagram contributions in Sec~\ref{sec:res}.}

In the large-$\beta_0$ limit of QCD~\cite{Beneke:1998ui}, one considers first the limit of large $n_f$, keeping $\alpha_s n_f \sim \mathcal{O}(1)$. The fermionic corrections to the gluon propagator acquire a special character since they count as $\mathcal{O}(1)$, and a dressed gluon propagator, known to all orders in $\alpha_s$, can be obtained. When no external gluons are present, such as in the case of this work, the leading-$n_f$ contributions to a given process can be obtained, to all orders, by replacing the gluon propagator in the calculation of the $\mathcal{O}(\alpha_s)$ QCD correction by the Borel transform of the dressed gluon propagator. The large-$\beta_0$ limit consists then in the replacement of the fermionic contribution to the QCD $\beta$ function, $\beta_{0f}$, proportional to $n_f$, by the complete one-loop QCD beta-function coefficient, $\beta_0$. This procedure, known as naive non-abelianization~\cite{Broadhurst:1994se,Beneke:1994qe}, effectively resums a set of non-Abelian graphs related to the running of the coupling, thereby restoring the non-Abelian character of the results.

Our calculation is performed with a modified version of the publicly available package {\tt MATAD}~\cite{Steinhauser:2000ry}, which is written in {\tt FORM}~\cite{Ruijl:2017dtg}. Working with the Borel-transformed dressed gluon propagator in the large-$\beta_0$ limit amounts, essentially, to a perturbative calculation with an analytically regularized gluon propagator. This requires that the exponent of the squared momentum in the gluon propagator be generalized to a real number~\cite{Beneke:1998ui}.
This was implemented through modifications of the original source code of {\tt MATAD}, since this generalization is not supported by the standard version.

The importance of calculations in the large-$\beta_0$ limit should not be overstated. Albeit realistic, it is a toy model for higher orders, but one that can lead to insights about the physics of renormalons, non-perturbative corrections, and even, in some cases, allow for estimates of unknown higher-order contributions. The fact that the leading-$n_f$ terms are exactly obtained can also be useful as an independent cross-check of future calculations.

We present, in analytical form, the Borel transform of the decay amplitude in the large-$\beta_0$ limit. The amplitude is computed as an expansion in the heavy-top limit and we provide the first two terms of the corresponding series in $\tau_t$. From this result, it is simple to extract in analytical form the exact leading-$n_f$ contributions to the QCD series to all orders in $\alpha_s$, with $n_f$ being the number of massless quarks. We reproduce the known leading-$n_f$ results up to 5 loops, and give for the first time the explicit expression at 6 and 7 loops. The large-$\beta_0$ result allows for a discussion of the different renormalon singularities, and we show that the leading ultraviolet (UV) renormalon is very prominent, leading to a sign alternating perturbative series (for not-so-large renormalization scale choices). The large-$\beta_0$ series reproduces the approximate magnitude of the known QCD perturbative coefficients, although with incorrect signs. This allows for a speculation about the size of yet unknown contributions to the QCD series.

This work is organized as follows. In Sec.~\ref{sec:theory} we set up the theoretical framework and the notation. Our main results are given in Sec.~\ref{sec:res} together with the discussion of some of their potential implications. In Sec.~\ref{sec:conclu} we present our conclusions. We relegate to App.~\ref{app:matad} a brief description of the modifications to {\tt MATAD} that are required to work in the large-$\beta_0$ limit, while in App. B we present the full results for the leading-$n_f$ coefficients, up to $\mathcal{O}(\tau_t)$, at 6 and 7 loops.

\section{Theoretical framework}
\label{sec:theory}

The decay width of the process $H \rightarrow \gamma \gamma$ starts at one-loop order in the SM and can be written in terms of bosonic and fermionic contributions as
\begin{equation}\label{eq:GammaHgagaDef}
\Gamma(H \rightarrow \gamma \gamma) =
\frac{M_H^3}{64 \pi} \mid A_W(\tau_W) + \sum_f A_f (\tau_f) \mid^2,
\end{equation}
where the first amplitude, $A_W$, is due to $W$ boson diagrams, while $A_f$ stems from the decay mediated by charged fermions with mass $m_f$~\cite{Djouadi:2005gi}. In the above expression, $\tau_f = \frac{M_H^2}{4m_f^2}$ and $\tau_W = \frac{M_H^2}{4M_W^2}$, where $m_W$ is the $W$-boson mass. The fermionic contributions are strongly dominated by the top-quark loop, with small subleading components due to the bottom quark and the tau lepton. Here, we focus exclusively on the top-quark contribution, $A_t$.
We work in the heavy-top limit and compute $A_t$ as an expansion around $\tau_t\to 0$, keeping the first two terms of the corresponding series.
We expect the heavy-top limit (i.e. the leading term) to be sufficient to estimate effects due to higher orders in perturbation theory, since the first $\tau_t$ correction is of about 10\%.
We compute the first subleading term to crosscheck this assumption.

We write the amplitude $A_t$ as an expansion in powers of the strong coupling, $\alpha_s$, as
\begin{equation}\label{eq:Atexp}
A_t = \hat{A}_t \sum_{n=0}^\infty A_t^{(n)} \bigg(\frac{\alpha_s}{\pi} \bigg)^n ,
\end{equation}
where
\begin{equation}\label{eq:LOnorm}
\hat{A}_t = N_c \frac{2Q_t^2 \alpha}{3 \pi v},
\end{equation}
with $\alpha$ being the fine-structure constant, $Q_t$ is the electromagnetic top charge, $N_c=3$ is the number of colors, and $v=1/\sqrt{\sqrt{2}G_F}$ is the vacuum expectation value of the Higgs field with the Fermi constant as input. With this convention (which is the same as Ref.~\cite{Sturm:2014nva}, for example), the leading order contribution to the amplitude $A_t$, Eq.~(\ref{eq:Atexp}), reads, after the expansion in $\tau_t$,
\begin{equation}
A_t^{(0)} = 1 + \frac{7}{30}\tau_t + \frac{2}{21}\tau_t^2 + \frac{26}{525}\tau_t^3 + \frac{512}{17325}\tau_t^4 + \frac{1216}{63063}\tau_t^5 + \mathcal{O}(\tau_t^6).
\end{equation}
Using expansion by regions (see for example Ref.~\cite{Smirnov:2002pj}), this series can be obtained from the computation of the loop integrals in the hard region, which leads to so-called massive tadpole integrals.
This remains true at higher orders for non-singlet diagrams.

\begin{figure}[t!]
\centering
\includegraphics[width=0.9\textwidth]{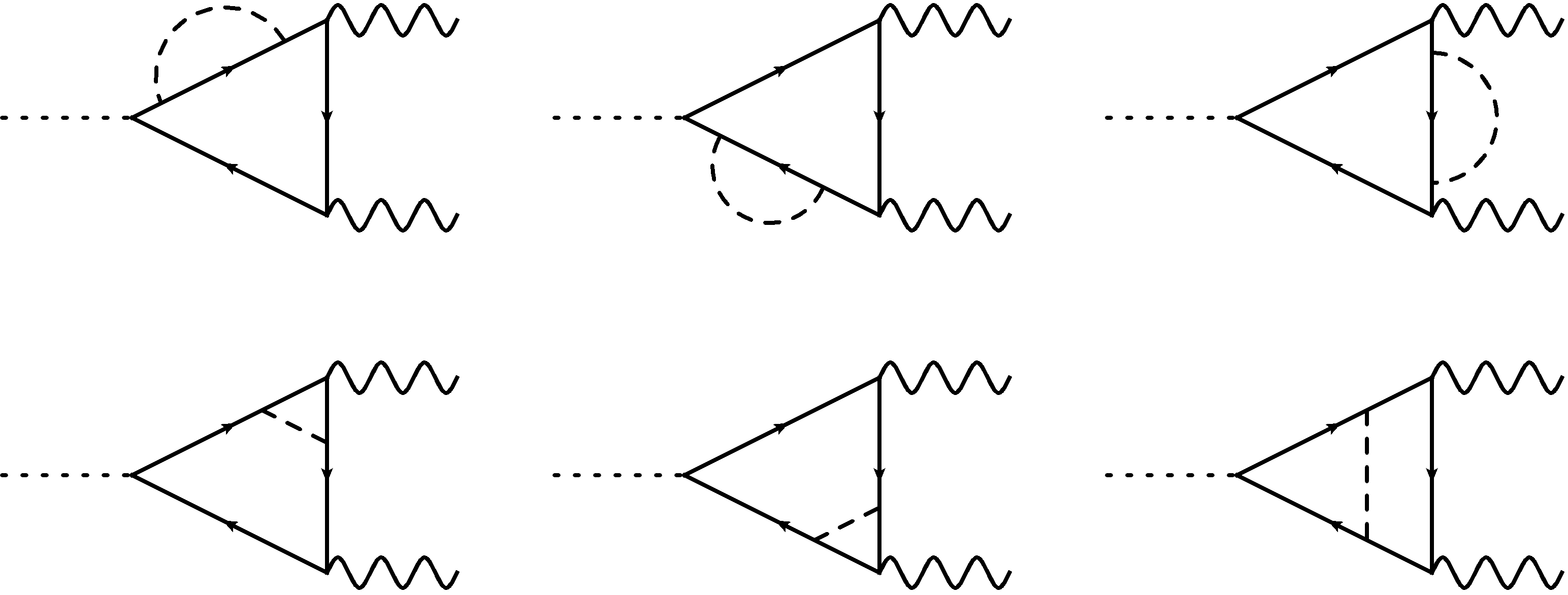}
\caption{Sample of diagrams for the calculation of the leading-$n_f$ terms for the decay $H \rightarrow \gamma \gamma$. The internal dashed lines represent the resummed gluon propagator, Eq~(\ref{eq:borel-gluon-prop}). }
\label{fig:Hgaga_nlo_sample_modified}
\end{figure}

The NLO term due to the top loop is obtained through the exchange of a virtual gluon between the internal top propagators~\cite{Djouadi:2005gi}. The topology of the main diagrams for the NLO process is shown in Fig.~\ref{fig:Hgaga_nlo_sample_modified}. There are a total of 12 diagrams at the two-loop level, and the final result for the amplitude up to corrections of $\mathcal{O}(\tau_t)$ is
\begin{equation}\label{eq:NLOcoeff}
A_{t}^{(1)} = C_F\left[ -\frac{3}{4}+ \tau_t\left( \frac{19}{90}-\frac{7}{20}\ell_\mu \right) +\mathcal{O}(\tau_t^2)\right] = -1 +\tau_t\left(\frac{38}{135}-\frac{7}{15}\ell_\mu \right) +\mathcal{O}(\tau_t^2)
\end{equation}
with $C_F = (N_c^2-1)/(2N_c)=4/3$, $\ell_\mu \equiv \ln \left( \frac{\mu^2}{\overline m_t^2} \right)$ and $\overline m_t\equiv \overline m_t(\mu)$ is the $\overline{\rm MS}$ top-quark mass.
Here, we have reproduced the NLO calculation using {\tt qgraf}\cite{Nogueira:1991ex} and {\tt MATAD}~\cite{Steinhauser:2000ry}, since this forms the basis for the computation in the large-$\beta_0$ limit, as we discuss further below.
Analytical expressions for higher order corrections, $A_t^{(2)}$ and $A_t^{(3)}$, up to four loops, can be found in Refs.~\cite{Sturm:2014nva,Davies:2021zbx}.

We turn now to the setup of the large-$\beta_0$ limit calculation. In the large-$\beta_0$ limit we deal with the perturbative series to all orders in $\alpha_s$. As is well known, in QCD, series of this type have coefficients that diverge factorially and the perturbative series is, at best, asymptotic. In this context it is convenient to work with the Borel transform of the series, which suppresses the factorial divergence of the coefficients and can have a finite radius of convergence. For the perturbative expansion of a quantity $R$ starting at $\mathcal{O}(\alpha_s)$ (without any loss of generality),
\begin{equation}
R \sim \sum_{n=0}^\infty r_n \alpha_s^{n+1},
\end{equation}
we define its Borel transform in the following way~\cite{Beneke:1998ui}
\begin{equation}
\label{eq:borel-transform-def}
B[R](t) = \sum_{n=0}^\infty r_n \frac{t^n}{n!}.
\end{equation}
The Borel transform is the inverse Laplace transform of the original series. The procedure can be inverted and a `true value' of the asymptotic series can be assigned (in the Borel sense) through the Laplace transform as
\begin{equation}
\label{eq:BorelInt}
R(\alpha_s)= \int_0^\infty {\rm d}t\, e^{-t/\alpha_s} B[R](t),
\end{equation}
provided the integral exists. In the Borel $t$-plane, singularities known as renormalons appear. They arise from infra-red (IR) and ultra-violet (UV) regions in the loop sub-graphs. The IR renormalons are particularly important here, since they appear on the right-hand side and obstruct the integration in Eq.~(\ref{eq:BorelInt}). Circumventing these singularities generates an imaginary ambiguity in the Borel integral. This ambiguity is related to non-perturbative terms from condensate matrix elements in the operator product expansion (OPE). Here, since the typical scale of the problem is quite high compared with $\Lambda_{\rm QCD}$, these OPE corrections --- and accordingly the imaginary ambiguities of IR renormalons --- are tiny and can, for all practical purposes, be neglected.

Using the definition in Eq.~(\ref{eq:borel-transform-def}), the dressed gluon propagator with a four-momentum $k$ obtained after resumming the massless-quark bubble-loop corrections in Borel space reads~\cite{Beneke:1998ui}
\begin{equation}\label{eq:borel-gluon-prop}
B[\alpha_s G_{\mu \nu}](u) = \frac{(-i)}{k^2} \bigg( g_{\mu \nu} - \frac{k_\mu k_\nu}{k^2} \bigg) \bigg( -\frac{\mu^2}{k^2}e^{-C} \bigg)^u + (-i) \xi \frac{k_\mu k_\nu}{k^4},
\end{equation}
where we introduced the variable $u$, defined as
\begin{equation}\label{eq:u-def}
u = -\beta_{0,f} t.
\end{equation}
(We recall that $\beta_{0f}$ are the fermionic contributions to $\beta_0$.) In Eq.~(\ref{eq:borel-gluon-prop}), $\mu^2$ is the renormalization scale and $\xi$ is the gauge parameter. The constant $C$ sets the renormalization scheme: in the $\overline{\text{MS}}$ scheme, $C = -5/3$; whereas in the MS scheme, for instance, $C = -5/3 + \gamma_E - \ln 4 \pi$.
Notice that, in the definition of the transformation, Eq.~(\ref{eq:borel-gluon-prop}), we multiplied the gluon propagator by $\alpha_s$; thus, the lowest order term in the $u$ expansion corresponds already to the NLO QCD correction.
In Eq.~(\ref{eq:borel-gluon-prop}), the factor $(-\mu^2 e^{-C})^u$ ensures scheme and scale invariance of the Borel integral result, Eq.~(\ref{eq:BorelInt}). We use the following definition for the QCD $\beta$ function and its coefficients:
\begin{equation}
\beta(\alpha_s)=\beta_0\alpha_s^2 +\beta_1\alpha_s^3 +\cdots=\mu^2\frac{{{\rm d}\,\alpha_s}}{{\rm d} \mu^2},
\end{equation}
with
\begin{equation}\label{eq:betaQCD}
\beta_0 = -\frac{11\,C_A}{12\pi} + \frac{n_f T_F}{3\pi},
\end{equation}
where $C_A=N_c=3$, $T_F=1/2$, and $n_f$ is the number of massless quark flavors. The second contribution on the right-hand side of Eq.~(\ref{eq:betaQCD}) is the fermionic contribution, $\beta_{0,f}=n_f/(6\pi)$.

Working in Landau gauge, with $\xi=0$, the Borel transform of the dressed propagator is essentially the original propagator with a modification in the exponent of the denominator momentum, $k^2 \rightarrow (k^2)^{1+u}$ --- which amounts to working with an analytically regularized gluon propagator. In our case, using Eq.~(\ref{eq:borel-gluon-prop}) as the gluon propagator in the NLO QCD calculation is sufficient to generate the exact leading-$n_f$ terms to all orders in $\alpha_s$. This is depicted in Fig.~\ref{fig:Hgaga_nlo_sample_modified}, where the main topologies involved in the diagrammatic calculation are shown and the dashed line represents the resummed, dressed, gluon propagator. The calculation of the diagrams of Fig.~\ref{fig:Hgaga_nlo_sample_modified} yields the main result of this paper. We implemented this calculation in \texttt{MATAD} with suitable modifications to the source code to account, essentially, for the new gluon propagator. Some technical details of this implementation can be found in Appendix~\ref{app:matad}. The source is also publicly available on GitHub.\footnote{The source files can be found in the following link \url{https://github.com/g-neves/hgg-large-beta0/tree/master/c_hgagalargeb0/hgagalargeb0}} In the next section we discuss the final results.

\section{Results}
\label{sec:res}

The Borel transform of the amplitude $A_t$ in the large-$\beta_0$ limit, up to $\mathcal{O}(\tau_t)$, denoted $B[A_{t,\text{large-}\beta_0}]$, in closed form, reads
\begin{align}
\hspace{-0.2cm}B[A_{t,\text{large-}\beta_0}] &= \nonumber
\hat A_t \frac{3 C_F}{4\pi}
\left( \frac{\mu^2}{\overline{m}_t^2} \right)^u \frac{e^{5u/3} (u^2-1) \Gamma(1-u) \Gamma(1+u)^3}{(1+2u) \Gamma(1+2u)} \nonumber\\
& \times \bigg\{ 1
+ \tau_t \bigg[
\frac{P(u)}{90 u (3 + 2 u)}
+ \frac{14 (1 + 2 u) \Gamma(1 - 2 u) \Gamma(1 + 2 u)}{15 (u^2-1)(2 - u) u \Gamma(1 - u)^2 \Gamma(1 + u)^2}
\bigg] \bigg\} \nonumber \\
& + \hat A_t \frac{3 C_F}{4\pi}
\left(-\frac{7}{45}\right) \tau_t \left[
\left( \frac{\mu^2}{\overline{m}_t^2} \right)^u e^{5u/3}
\frac{6 (1 - u) \Gamma(u) \Gamma(1 - 2 u)}{\Gamma(3 - u)}
+ \frac{\tilde{G}_0(u)}{u} \right],
\label{eq:borel-A_t}
\end{align}
where $\hat A_t$ is defined in Eq.~(\ref{eq:LOnorm})
and $P(u)= 126 + 155 u + 180 u^2 + 100 u^3 + 9 u^4$.
The last line in Eq.~(\ref{eq:borel-A_t}) stems from the relation between the $\overline{\rm MS}$ and pole mass in the large-$\beta_0$ limit~\cite{Beneke:1994sw,Ball:1995ni}.
Removing this line effectively converts the mass scheme from the former to the latter.
The function $\tilde{G}_0$ is defined as a power series in $u$ with coefficients $g_n/n!$, where the $g_n$ are obtained from the generating function
\[
G_0(u)= -\frac{(3 + 2 u) \Gamma(4 + 2 u)}{3 \Gamma(1 - u) \Gamma(2 + u)^2 \Gamma(3 + u)} \,.
\]
The expression in Eq.~(\ref{eq:borel-A_t}) exhibits the renormalon singularities arising from UV and IR regions of loop sub-graphs, which are encoded in the $\Gamma$ functions. In the leading term, corresponding to an infinitely heavy top quark, the UV renormalons, which are located at negative integer values of the variable $u$, are all double poles, with the sole exception of the leading UV pole at $u=-1$, which is simple. This happens due to a partial cancellation with the $(u^2-1)/\Gamma(1+2u)$ term. Note also that there is no singularity at $u=-1/2$ and the function is regular at this point, as expected. The IR renormalons, on the other hand, are all simple poles with the leading singularity at $u=2$ being a quartic IR sensitivity, connected with dimension four corrections in the Operator Product Expansion. Regarding the dominant renormalon singularities another observation is that the residue of the leading UV pole and that of the leading IR pole are of the same order. Their ratio being of order 1 implies that the UV renormalon, being the closest to the origin, should take over rather earlier and lead to a sign alternating series. This, however, depends on the renormalization scale that is chosen: larger values of $\mu$ enhance the IR poles and suppress the UV contributions, which postpones the dominance of the UV renormalon, reducing its residue, and consequently delays the sign alternation. The structure of the term proportional to $\tau_t$ is similar. The main difference here is the appearence of a term without the prefactor $\mu^{2u}e^{5u/3}$, which ensures scheme and scale invariance of the Borel integral. This happens because the last term in Eq.~(\ref{eq:borel-A_t}) is a consequence of the renormalization scheme (and scale) dependence of the quark mass. Finally, we note that the expression is regular at $u=0$.

From the Taylor expansion of Eq.~(\ref{eq:borel-A_t}) one recovers the perturbative expansion in $\alpha_s$ in the large-$\beta_0$ limit. Of course, the leading term must correspond to the exact NLO QCD result, and it does lead to $A_t^{(1)}$ of Eq.~(\ref{eq:NLOcoeff}), as expected. Then, further expanding in $u$ and performing the substitution $u = -\beta_{0,f} t$, one can re-obtain the leading-$n_f$ contributions, i.e. the coefficients of terms proportional to $\alpha_s^n n_f^{n-1}$, that we denote $A_{t,{n_f^{n-1}}}^{(n)}$ of the exactly known QCD corrections in perturbation theory up to $\mathcal{O}(\tau_t)$. At 3 and 4 loops we get
\begin{align}\label{eq:At2nf}
A_{t, n_f}^{(2)} =& -C_F T_F n_f \left[ \frac{1}{12} - \frac{\ell_\mu}{4} - \tau_t\left(\frac{1213}{4320} +\frac{29 \ell_\mu}{1080}+\frac{7 \ell_\mu^2}{120}\right) \right], \\\label{eq:At3nf}
A_{t, n_f^2}^{(3)} =& -C_F T_F^2 n_f^2 \Bigg[ \frac{19}{108} - \frac{\ell_\mu}{18} + \frac{\ell_\mu^2}{12} \nonumber \\
&+\tau_t \left(-\frac{8657}{38880}+\frac{49 \zeta_3}{270}+\frac{121 \ell_\mu}{810}+\frac{29 \ell_\mu^2}{3240}+\frac{7\ell_\mu^3}{540} \right) \Bigg],
\end{align}
where $\zeta_i$ is the Riemann Zeta Function evaluated at $i$ and we recall that $\ell_\mu \equiv \ln \left( \frac{\mu^2}{\overline m_t^2} \right)$. These results are in agreement with Refs.~\cite{Sturm:2014nva, Maierhofer:2012vv}.
At 5 loops, we reproduce the result for the leading-$n_f$ term in the heavy-top limit~\cite{Sturm:2014nva} and obtain for the first time the leading-$n_f$ terms of the $\mathcal{O}(\tau_t)$ correction as
\begin{align}\label{eq:At4nf}
&A_{t, n_f^3}^{(4)} = -C_F T_F^3 n_f^3 \Bigg[ \frac{487}{972} - \frac{\zeta_3}{3} - \frac{19}{108}\ell_\mu + \frac{\ell_\mu^2}{36} - \frac{\ell_\mu^3}{36}\nonumber \\
&+\tau_t\left( -\frac{2873063}{8398080}+\frac{7 \pi ^4}{2592}-\frac{89 \zeta_3}{2160}+\ell_\mu \left(\frac{673}{3240}-\frac{7 \zeta_3}{45}\right)-\frac{121 \ell_\mu^2}{1620}-\frac{29 \ell_\mu^3}{9720}-\frac{7
\ell_\mu^4}{2160} \right)\bigg].
\end{align}
A new result of this work is the leading $n_f$-terms to all orders in perturbation theory. As an example, in the heavy-top limit, the N$^5$LO leading-$n_f$ coefficient, $A_{t, n_f^4}^{(5)}$, reads
\begin{align}\label{eq:At5nf}
A_{t, n_f^4}^{(5)} = -C_F T_F^4 n_f^4 \bigg[ \frac{9613}{8748} - \frac{\pi^4}{135} - \frac{4 \zeta_3}{27} + \ell_\mu \left( \frac{4 \zeta_3}{9} - \frac{487}{729} \right)
+ \frac{19}{162} \ell_\mu^2 - \frac{\ell_\mu^3}{81} + \frac{\ell_\mu^4}{108} \bigg],
\end{align}
while the N$^6$LO leading-$n_f$ coefficient, $A_{t, n_f^5}^{(6)}$ is
\begin{align}\label{eq:At6nf}
A_{t, n_f^5}^{(6)} = -C_F T_F^5 n_f^5 \bigg[ & \frac{307765}{78732} - \frac{ \pi^4}{243} - \frac{190}{243} \zeta_3 - \frac{20}{9} \zeta_5
+ \ell_\mu \bigg( - \frac{48065}{26244} + \frac{\pi^4}{81} + \frac{20}{81} \zeta_3 \bigg) \nonumber\\
& + \ell_\mu^2 \bigg(\frac{2435}{4374} - \frac{10}{27} \zeta_3 \bigg) - \frac{95}{1458} \ell_\mu^3 + \frac{5}{972} \ell_\mu^4 - \frac{1}{324} \ell_\mu^5 \bigg].
\end{align}
The corresponding $\mathcal{O}(\tau_t)$ corrections are lengthy and are given in App.~\ref{app:tauterms}.
The higher-order coefficients can be easily generated from Eq.~(\ref{eq:borel-A_t}) following the procedure outlined above.

\begin{figure}[!t]
\centering
\subfloat[]{\label{fig:pt-series}{\includegraphics[width=0.5\textwidth]{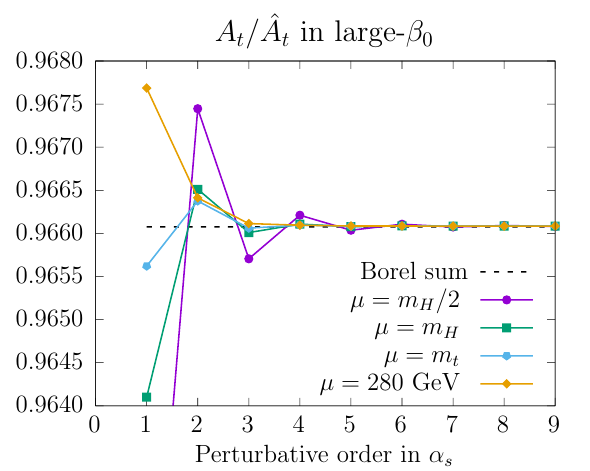}}}\hfill
\subfloat[]{\label{fig:scale}{\includegraphics[width=0.5\textwidth]{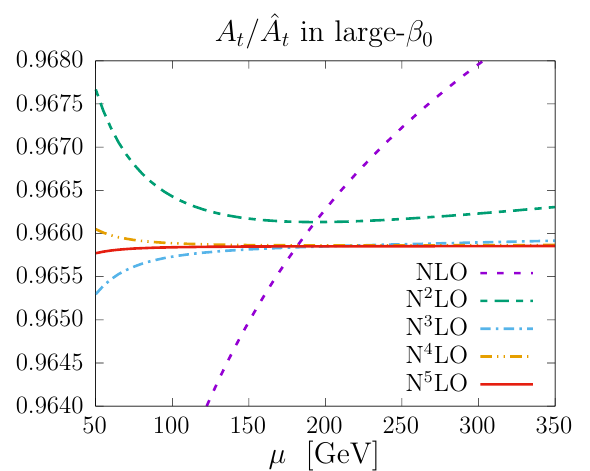}}}
\caption{(a) Perturbative series, Eq.~(\ref{eq:Atexp}), in the large-$\beta_0$ limit for four different values of the renormalization scale $\mu$, order by order. The Borel sum of the series (see text) is shown as the horizontal dashed line. (b) Perturbative series as a function of the renormalization scale up to N$^5$LO (or $\mathcal{O}(\alpha_s^5)$). The scale variation range $50~{\rm GeV} \leq \mu \leq 350~{\rm GeV}$ contains the interval $m_H/2\leq \mu\leq 2m_t$.}
\end{figure}

Let us now discuss the series in the large-$\beta_0$ limit, after naive non-abelianization, with the replacement $\beta_{0f}\to \beta_0$. This transformation preserves the leading $n_f$ coefficients $A_{t,{n_f^{n-1}}}^{(n)}$ and generates approximate results for all sub-leading powers of $n_f$. The perturbative series thus obtained is, at best, a good qualitative approximation to the full QCD results. With this caveat in mind, we show in Fig.~\ref{fig:pt-series} the perturbative series for four different choices of the renormalization scale $\mu$, order by order in perturbation theory. The horizontal line is the Borel sum of the series, given by the integral of Eq.~(\ref{eq:BorelInt}).\footnote{The series, strictly speaking, is not Borel summable due to the IR renormalons, which obstruct the integration. A prescription to circumvent them must be chosen, which generates an imaginary ambiguity which scales as $e^{-p/\alpha_s}$ (where $p>0$), related to the size of non-perturbative corrections. However, these contributions, at the Higgs mass scale, are very small, which leads to only a tiny ambiguity which is not visible in our plots and can be ignored here.} In Fig.~\ref{fig:scale} we display the scale dependence of the result, order by order, varying the renormalization scale $\mu$ in the logarithms $\ell_\mu$ (see Eqs.~(\ref{eq:At2nf})-(\ref{eq:At6nf})) and in $\alpha_s^{(n_f=5)}(\mu)$. The running of $\alpha_s^{(n_f=5)}(\mu)$ is performed at one loop, for consistency with the large-$\beta_0$ limit, using as input $\alpha_s^{(n_f=5)}(m_Z)=0.1179$ while for the top mass the value of $\overline m_t(\overline m_{\rm t})=163.643$~GeV~\cite{Beneke:2021lkq} is kept fixed, since the $\overline{\rm MS}$ quark-mass running is of order $1/\beta_0$ itself, which would then generate $1/\beta_0^2$ terms not included in the perturbative coefficients~\cite{Grozin:2003gf,Boito:2021wbj}. The stabilization with respect to scale variations is clearly observed when the order is increased, starting from N$^4$LO, or $\mathcal{O}(\alpha_s^4)$, with the N$^5$LO result being extremely stable and in excellent agreement with the Borel sum.

Before comparing with the QCD calculation, an observation regarding the singlet diagrams is in order. In Ref.~\cite{Sturm:2014nva}, it is shown that with the scale choice $\mu=\overline m_t(\overline m_t)$ (which resums the logarithms $\ell_\mu$) at the 3-loop level the singlet diagrams are approximately of the same magnitude as the non-singlet diagrams, while at a lower scale, $\mu=m_H$, the singlet contribution is significantly enhanced and is approximately three times larger than the non-singlet contribution. It is argued that one might expect the same to happen at 4- and 5-loop orders. We have checked that this is generally the case at the 4-loop level for
the singlet diagrams where both photons couple to massless quarks, using the result from Ref. [17]. However, here the singlet contribution is suppressed at $\mu=\overline{m}_t(\overline{m}_t)$, before dominating around $\mu\approx 270$~GeV (due to the non-singlet contribution becoming zero) and becoming relatively smaller again at even higher scales. Thus, if the large-$\beta_0$ limit represents well quantitatively the non-singlet diagrams in full QCD, at higher values of $\mu$, the difference between the result in QCD and in large-$\beta_0$ should decrease. Below, the results we discuss are for $\mu=\overline{m}_t(\overline{m}_t)$, which reduces the singlet contributions.

After naive nonabelianization, our result for $A_{t,\text{large-}\beta_0}$ at the scale $\mu=\overline m_t(\overline m_t)$ reads ($a_s\equiv \alpha_s(\overline m_t(\overline m_t))/\pi$)
\begin{align}
\frac{A_{t,\text{large-}\beta_0}}{\hat A_t}
=& 1 + 0.0342 \tau_t
- ( 1 - 0.0412 \tau_t )\, a_s
+ ( 0.6389 - 0.3153 \tau_t )\, a_s^2
\notag \\
& - ( 7.755 - 0.0291 \tau_t )\, a_s^3
+ ( 25.43 - 4.773 \tau_t )\, a_s^4
- ( 290.4 - 16.78 \tau_t )\, a_s^5
\notag \\
& + ( 2212. - 277.8 \tau_t )\, a_s^6+\cdots
\label{eq:AtLb0}
\end{align}
This result shows that in the large-$\beta_0$ limit the pattern of sign alternation is already present, systematically, starting at $\mathcal{O}(\alpha_s^2)$. This is a consequence of the dominant behavior of the leading UV singularity. (This pattern disappears if even larger values of $\mu$ are used. For example, for $\mu=280$~GeV the coefficients have fixed signs, as can be seen in Fig.~\ref{fig:pt-series} for the leading term in $\tau_t$.)
As expected, we find that the subleading terms in $\tau_t$ are generally about an order of magnitude smaller than the leading ones (the $\mathcal{O}(\alpha_s^2)$ coefficient is an exception).

We can now compare our result for the leading term in $\tau_t$ with results from Refs.~\cite{Sturm:2014nva,Davies:2021zbx}.
In the same conventions, and again for $\mu=\overline m_t(\overline m_t)$ and $a_s=\alpha_s^{(n_f=5)}(\overline m_t(\overline m_t))/\pi$, using $m_H=125.25$~GeV, we find for the QCD result
\begin{align}\label{eq:AtQCD}
\frac{A_{t}}{\hat A_t}
= 1 - a_s &- (1.167+(1.020-1.440i)_{\rm si})\,a_s^2 \nonumber \\&+(6.669+(0.7716+16.49\,i)_{\rm si})\,a_s^3 -(22.31+c_4)\,a_s^4+\cdots
\end{align}
At $\mathcal{O}(a_s^4)$ only the contributions where the photons couple to a massive top-quark loop are known from Ref.~\cite{Sturm:2014nva} (we have re-expressed the result of Ref.~\cite{Sturm:2014nva} in terms of $a_s=\alpha_s^{(n_f=5)}/\pi$) and
the constant $c_4$ has yet to be calculated. But these contributions do dominate the real part of the $\mathcal{O}(\alpha_s^3)$ coefficient, as can be seen comparing the results of Refs.~\cite{Sturm:2014nva,Davies:2021zbx}. The singlet contributions where the photons couple to massless quarks at $\mathcal{O}(a_s^2)$ and $\mathcal{O}(a_s^3)$ are indicated by `si'. The imaginary part in the singlet contributions is required by unitarity. In the decay width, however, the first contribution arising from an imaginary part, namely from ${\rm Im}(A_t^{(2)})$, starts at $\mathcal{O}(\alpha_s^4)$ and can be expected to be suppressed, since it competes with the real parts from higher-order coefficients, which tend to be significantly larger, as seen in the QCD result. We therefore compare the real parts of the coefficients of Eqs.~(\ref{eq:AtLb0}) and~(\ref{eq:AtQCD}). From these results we see that the magnitude of the coefficients obtained in the large-$\beta_0$ limit are not too far from the magnitude of the known QCD coefficients, but the sign is wrong in all cases. In QCD, the systematic sign alternation is not yet established at order $\mathcal{O}(\alpha_s^2)$ although the signs do alternate starting from $\mathcal{O}(\alpha_s^3)$ for the known terms. This is an indication that the UV renormalon in QCD should be less prominent than in the large-$\beta_0$ limit, in agreement with other processes where the large-$\beta_0$ result is known~\cite{Boito:2018rwt, Boito:2021wbj,Benitez-Rathgeb:2022yqb}.

Based on the observation that the large-$\beta_0$ result roughly reproduces the size (but not the sign) of the QCD contributions, we can speculate that the $\alpha_s^5$ non-singlet coefficient would have the same magnitude as the one shown in Eq.~(\ref{eq:AtLb0}). Given the important contribution of the singlet diagrams, it is safer to attach a 100\% uncertainty to this result, which leads to the following estimate for the $\mathcal{O}(a_s^5)$ contribution to $A_t$ for $\mu=\overline m_t(\overline m_t)$:
\begin{equation}
A_t^{(5)}\approx 300 \pm 300.
\end{equation}
Using this value and the QCD series of Eq.~(\ref{eq:AtQCD}), in the decay width, this implies a correction of,
\begin{equation}
\Gamma(H\to\gamma\gamma)\Big|_{\alpha_s^5}-\Gamma(H\to\gamma\gamma)\Big|_{\alpha_s^4} \approx -0.00014 \,{\rm keV}, \qquad [\mu=\overline m_t(\overline m_t)],
\end{equation}
which is about as large as the contribution from the partial $\alpha_s^4$ correction calculated in Ref.~\cite{Sturm:2014nva}, which amounted to mere 0.02\textperthousand~of the total decay width. (For this estimate we have added the LO $W$ contribution and top-mass corrections in the 1- and 2-loop QCD results --- see, e.g., the expressions given in Ref.~\cite{Maierhofer:2012vv}.)
If significantly lower renormalization scales are used one finds, for $\mu=m_H/2$ for example,
\begin{equation}
\Gamma(H\to\gamma\gamma)\Big|_{\alpha_s^5}-\Gamma(H\to\gamma\gamma)\Big|_{\alpha_s^4} \approx -0.00029 \,{\rm keV}, \qquad [\mu= m_H/2],
\end{equation}
which is, at this scale, 2.5 times smaller than the (partial) $\alpha_s^4$ correction. In all cases, we find that the $\mathcal{O}(\alpha_s^5)$ correction is at most as large as the $\mathcal{O}(\alpha_s^4)$ contribution, which means that estimating the truncation error from the last included term is safe.
This clearly indicates that the QCD contributions are under very good control here, with an uncertainty from the truncation which is an order of magnitude smaller than the uncertainty arising parametrically from the value of $\alpha_s$ itself, for example, and much smaller than the uncertainty from $m_H$ which amounts to about $0.04$~keV.

\section{Conclusions}
\label{sec:conclu}

In this work we presented the result for the QCD corrections to $H\to\gamma\gamma$ in the large-$\beta_0$ limit, providing the first two terms in the heavy-top expansion. The analytical result was obtained with a modified version of \texttt{MATAD}~\cite{Steinhauser:2000ry}. From the Borel transformed amplitude, upon re-expanding in the Borel variable, we can reconstruct the perturbative series to all orders in $\alpha_s$ in the large-$\beta_0$ limit. In particular, the exact analytical results for the leading-$n_f$ contribution at each order can be extracted and, apart from reproducing the known results, we have given explicit expressions for the previously unknown contributions at 6 and 7 loops, in Eqs.~(\ref{eq:At5nf}) and~(\ref{eq:At6nf}). The higher-order leading-$n_f$ coefficients can easily be extracted from the analytical result for the Borel transformed amplitude, Eq.~(\ref{eq:borel-A_t}). These results can serve as a partial cross-check for future calculations, should the full QCD corrections be independently calculated by other groups.

Furthermore, the exact knowledge of the Borel transform in the large-$\beta_0$ limit allows for a study of the renormalon singularities. We find the usual and expected towers of UV and IR renormalons: the leading UV renormalon, located at $u=-1$ in the Borel plane, being the closest to the origin and dominates the series at high orders, while the leading IR renormalon is related to a quartic IR sensitivity, and appears at $u=2$. In the large-$\beta_0$ limit, the UV renormalon has a somewhat large residue and it dominates the series for values of the renormalization scale below 200~GeV or so. The perturbative series has a systematic sign alternation which is not observed in QCD. In QCD, the coefficients can have different signs but no systematic alternation is observed, which indicates a weaker UV renormalon, as observed in other processes~\cite{Boito:2018rwt,Boito:2021wbj,Benitez-Rathgeb:2022yqb}, and a more complicated interplay between UV and IR contributions.

The perturbative series in the large-$\beta_0$ limit has coefficients of roughly the same order of magnitude as in full QCD up to $\mathcal{O}(\alpha_s^3)$, or even $\mathcal{O}(\alpha_s^4)$ if we compare with the partial results of Ref.~\cite{Sturm:2014nva}, although the signs of the coefficients are opposite. Assuming this observation survives at even higher orders, we estimate the six-loop coefficient to be $A_t^{(5)} \approx 300\pm 300$, for $\mu=\overline m_t(\overline m_t)$. This, in turn, leads to a contribution to the decay width of at most 0.00015~keV, which is significantly below the parametric uncertainty induced by $m_H$ and $\alpha_s$ and much smaller than the available estimates of the $\mathcal{O}(\alpha_s^4)$ contribution~\cite{Sturm:2014nva}.

\appendix
\section{Modifications to \texttt{MATAD}'s source code}
\label{app:matad}

In this appendix we briefly describe the technical implementation in {\tt MATAD} of the calculation of the diagrams of Fig.~\ref{fig:Hgaga_nlo_sample_modified}.
The implementation of the large-$\beta_0$ limit calculation in \texttt{MATAD} requires some modifications to the source code. The first modification deals with the Borel transformed dressed gluon propagator, where the variable $u$ of Eq.~(\ref{eq:borel-gluon-prop}), which is not supported in the original package, must be included. This basically generalizes the exponent of the $k^2$ term in the denominator of the gluon propagator to the real domain.
The second modification takes into account the inclusion of the variable $u$ into the 2-loop tadpole integrals, where the modified gluon propagator enters the calculation.

The first modification is to the gluon propagator to account for Eq.~(\ref{eq:borel-gluon-prop}). We arranged the diagrams such that the gluon propagators always carried momentum $p_3$, without any external momenta. With this, the modification is rather simple, and we only need to multiply the original propagator by a new function that we call \texttt{Denu}, which accounts for the factors of $u$ in the modified propagator.
In standard \texttt{FORM} notation, this amounts~to
\begin{lstlisting}
id Dg(?x) = Dg(?x)*Denu(u);
\end{lstlisting}

In the next step, one identifies the powers in the quark and gluon propagators and unifies the scalar integrals into a single function with the relevant coefficients in order to make contact with the analytic calculation of the relevant two-loop integrals. As discussed previously, $p_3$ was reserved for the gluon propagator; the momenta $p_1$ and $p_2$ were assigned to the top propagators. In the following expression, \texttt{s1m} and \texttt{s2m} represent the quark propagators with momentum $p_1$ and $p_2$, respectively. In \texttt{FORM} code, the unification of the scalar integrals into a function with the relevant coefficients reads
\begin{lstlisting}
id s1m^a1?*s2m^a2?/p3.p3^a3? = f(a1,a2,a3);
\end{lstlisting}
After this, we simply exclude the scaleless integrals which integrate to zero in dimensional regularization.

Since we are interested in the expansion in $\tau_t$ for non-singlet diagramsx, we need to consider only the hard region of the loop integrals. In this region the integrals become massive tadpoles. We can then perform the corresponding scalar loop integrals using the well-known formula
\begin{align}
&\int \frac{ \mathrm{d}^Dk\, \mathrm{d}^D\ell}{(-k^2 + m^2)^{a_1} (-\ell^2 + m^2)^{a_2} [-(k+\ell)^2]^{a_3}} \nonumber \\
&= (i\pi^{D/2})^2 \frac{\Gamma(a_1 + a_3 + \varepsilon -2) \Gamma(a_2 + a_3 + \varepsilon -2) \Gamma(2-\varepsilon -a_3)}{\Gamma(a_1) \Gamma(a_2) \Gamma(2-\varepsilon)} \nonumber \\
& \times \frac{\Gamma(a_1 + a_2 + a_3 + 2\varepsilon - 4)}{\Gamma(a_1 + a_2 + 2a_3 + 2\varepsilon - 4)} (m^{-2})^{a_1 + a_2 + a_3 + 2\varepsilon - 4}.
\end{align}
The $u$ variable is attached to the massless propagator, i.e., in our case it appears together with the $a_3$ variable. We implement this with the following \texttt{id} statement:
\begin{lstlisting}
id f(a1?,a2?,a3?)*Denu(u?) = eMu(u) * M^(2*(4 - a1-a2-a3))
* Gam(a1+a3-2,1,u)*Gam(a2+a3-2,1,u)*Gam(2-a3,-1,-u)
* iGam(a1,0,0)*iGam(a2,0,0)*iGam(2,-1,0)
* Gam(a1+a2+a3-4,2,u)*iGam(a1+a2+2*a3-4,2,2*u);
.sort
\end{lstlisting}
On the r.h.s. of the equality sign, \texttt{eMu} is the function containing the additional factors of the Borel transformed gluon propagator in Eq.~(\ref{eq:borel-gluon-prop}), that is
\begin{equation*}
\text{\texttt{eMu}(u)} \equiv \bigg( -\frac{\mu^2}{m_t^2}e^{-C} \bigg)^u.
\end{equation*}
The functions \texttt{Gam} and \texttt{iGam} (\texttt{iGam} $\equiv$ 1/\texttt{Gam}) are the $\Gamma$-function and its inverse using the following notation
\begin{equation}
\Gamma(a+b\epsilon+cu) \equiv \texttt{Gam(a,b,c)},
\end{equation}
and analogously for the inverse of the $\Gamma$ function.

\section{\boldmath Leading top-mass corrections in the large-\texorpdfstring{$\beta_0$}{beta0} limit}
\label{app:tauterms}

Here we provide the explicit results for the leading-$n_f$ terms up to $\mathcal{O}(\tau_t)$ at 6 and 7 loops, which were omitted in Eq.~(\ref{eq:At5nf}) and~(\ref{eq:At6nf}) for the sake of brevity. The complete expressions are given below:
\begin{align}
A_{t, n_f^4}^{(5)} &= -C_F T_F^4 n_f^4 \Bigg[ \frac{9613}{8748} - \frac{\pi^4}{135} - \frac{4 \zeta_3}{27} + \ell_\mu \left( \frac{4 \zeta_3}{9} - \frac{487}{729} \right)
+ \frac{19}{162} \ell_\mu^2 - \frac{\ell_\mu^3}{81} + \frac{\ell_\mu^4}{108} \nonumber \\
&+\tau_t\Bigg( -\frac{70933639}{62985600}+\frac{11567 \zeta_3}{29160}+\frac{847 \zeta_5}{1350}-\frac{239 \pi ^4}{291600}\nonumber \\
&+\ell_\mu \left(\frac{177263}{393660}-\frac{7 \pi
^4}{2025}+\frac{58 \zeta_3}{1215}\right)+\ell_\mu^2 \left(-\frac{673}{4860}+\frac{14 \zeta_3}{135}\right)+\frac{121 \ell_\mu^3}{3645}\nonumber \\ &+\frac{29 \ell_\mu^4}{29160}+\frac{7
\ell_\mu^5}{8100} \Bigg)\Bigg],\\
A_{t, n_f^5}^{(6)} &= -C_F T_F^5 n_f^5 \Bigg[ \frac{307765}{78732} - \frac{ \pi^4}{243} - \frac{190}{243} \zeta_3 - \frac{20}{9} \zeta_5
+ \ell_\mu \bigg( - \frac{48065}{26244} + \frac{\pi^4}{81} + \frac{20}{81} \zeta_3 \bigg) \nonumber\\
& + \ell_\mu^2 \bigg(\frac{2435}{4374} - \frac{10}{27} \zeta_3 \bigg) - \frac{95}{1458} \ell_\mu^3 + \frac{5}{972} \ell_\mu^4 - \frac{1}{324} \ell_\mu^5 \nonumber \\
&+\tau_t \Bigg(
-\frac{4660780709}{1360488960}+\frac{69647 \pi ^4}{6298560}+\frac{721 \pi ^6}{393660}+\frac{968539
\zeta_3}{1049760}-\frac{5033 \zeta_3^2}{14580}-\frac{1399 \zeta_5}{5832}\nonumber \\
&+\ell_\mu
\left(\frac{885605}{472392}+\frac{29 \pi ^4}{21870}-\frac{484 \zeta_3}{729}-\frac{28 \zeta_5}{27}\right) +\ell_\mu^2
\left(-\frac{177263}{472392}+\frac{7 \pi ^4}{2430}-\frac{29 \zeta_3}{729}\right) \nonumber \\
&+\ell_\mu^3
\left(\frac{673}{8748}-\frac{14 \zeta_3}{243}\right)-\frac{121 \ell_\mu^4}{8748}-\frac{29 \ell_\mu^5}{87480}-\frac{7
\ell_\mu^6}{29160}
\Bigg)\Bigg].
\end{align}

\section*{Acknowledgments}
GdN thanks the Particle Physics Group of the Universit\"at Siegen, where part of this work was carried out, for hospitality.
DB's work was supported in part by the S\~ao Paulo Research Foundation (FAPESP) Grants
No.~2015/20689-9 and No.~2021/06756-6, by CNPq Grant No.\ 308979/2021-4. The work of GdN was suppported by FAPESP Grants No.~2018/12305-4 and No.~2019/17046-0.
The work of JP was supported by the Deutsche Forschungsgemeinschaft (DFG, German Research Foundation) under grant 396021762 -- TRR 257 ``Particle Physics Phenomenology after the Higgs Discovery.''

\bibliographystyle{jhep}
\bibliography{References}

\providecommand{\href}[2]{#2}\begingroup\raggedright\begin{thebibliography}{10}

\bibitem{Ellis:1975ap}
J.~R. Ellis, M.~K. Gaillard and D.~V. Nanopoulos, \emph{{A Phenomenological
  Profile of the Higgs Boson}},
  \href{https://doi.org/10.1016/0550-3213(76)90382-5}{\emph{Nucl. Phys. B}
  {\bfseries 106} (1976) 292}.

\bibitem{Shifman:1979eb}
M.~A. Shifman, A.~I. Vainshtein, M.~B. Voloshin and V.~I. Zakharov,
  \emph{{Low-Energy Theorems for Higgs Boson Couplings to Photons}},
  {\emph{Sov. J. Nucl. Phys.} {\bfseries 30} (1979) 711}.

\bibitem{Okun:1982ap}
L.~B. Okun, \emph{{Leptons and Quarks}: {Special Edition Commemorating the
  Discovery of the Higgs Boson}}. North-Holland, Amsterdam, Netherlands, 1982,
  \href{https://doi.org/10.1142/9162}{10.1142/9162}.

\bibitem{Gavela:1981ri}
M.~B. Gavela, G.~Girardi, C.~Malleville and P.~Sorba, \emph{{A Nonlinear R(xi)
  Gauge Condition for the Electroweak SU(2) X U(1) Model}},
  \href{https://doi.org/10.1016/0550-3213(81)90529-0}{\emph{Nucl. Phys. B}
  {\bfseries 193} (1981) 257}.

\bibitem{Djouadi:1990aj}
A.~Djouadi, M.~Spira, J.~J. van~der Bij and P.~M. Zerwas, \emph{{QCD
  corrections to gamma gamma decays of Higgs particles in the intermediate mass
  range}}, \href{https://doi.org/10.1016/0370-2693(91)90879-U}{\emph{Phys.
  Lett. B} {\bfseries 257} (1991) 187}.

\bibitem{Melnikov:1993tj}
K.~Melnikov and O.~I. Yakovlev, \emph{{Higgs $\to$ two photon decay: QCD
  radiative correction}},
  \href{https://doi.org/10.1016/0370-2693(93)90507-E}{\emph{Phys. Lett. B}
  {\bfseries 312} (1993) 179}
  [\href{https://arxiv.org/abs/hep-ph/9302281}{{\ttfamily hep-ph/9302281}}].

\bibitem{Djouadi:1993ji}
A.~Djouadi, M.~Spira and P.~M. Zerwas, \emph{{Two photon decay widths of Higgs
  particles}}, \href{https://doi.org/10.1016/0370-2693(93)90564-X}{\emph{Phys.
  Lett. B} {\bfseries 311} (1993) 255}
  [\href{https://arxiv.org/abs/hep-ph/9305335}{{\ttfamily hep-ph/9305335}}].

\bibitem{Inoue:1994jq}
M.~Inoue, R.~Najima, T.~Oka and J.~Saito, \emph{{QCD corrections to two photon
  decay of the Higgs boson and its reverse process}},
  \href{https://doi.org/10.1142/S0217732394001003}{\emph{Mod. Phys. Lett. A}
  {\bfseries 9} (1994) 1189}.

\bibitem{Spira:1995rr}
M.~Spira, A.~Djouadi, D.~Graudenz and P.~M. Zerwas, \emph{{Higgs boson
  production at the LHC}},
  \href{https://doi.org/10.1016/0550-3213(95)00379-7}{\emph{Nucl. Phys. B}
  {\bfseries 453} (1995) 17}
  [\href{https://arxiv.org/abs/hep-ph/9504378}{{\ttfamily hep-ph/9504378}}].

\bibitem{Fleischer:2004vb}
J.~Fleischer, O.~V. Tarasov and V.~O. Tarasov, \emph{{Analytical result for the
  two loop QCD correction to the decay H $\to$ 2 gamma}},
  \href{https://doi.org/10.1016/j.physletb.2004.01.063}{\emph{Phys. Lett. B}
  {\bfseries 584} (2004) 294}
  [\href{https://arxiv.org/abs/hep-ph/0401090}{{\ttfamily hep-ph/0401090}}].

\bibitem{Harlander:2005rq}
R.~Harlander and P.~Kant, \emph{{Higgs production and decay: Analytic results
  at next-to-leading order QCD}},
  \href{https://doi.org/10.1088/1126-6708/2005/12/015}{\emph{JHEP} {\bfseries
  12} (2005) 015} [\href{https://arxiv.org/abs/hep-ph/0509189}{{\ttfamily
  hep-ph/0509189}}].

\bibitem{Aglietti:2006tp}
U.~Aglietti, R.~Bonciani, G.~Degrassi and A.~Vicini, \emph{{Analytic Results
  for Virtual QCD Corrections to Higgs Production and Decay}},
  \href{https://doi.org/10.1088/1126-6708/2007/01/021}{\emph{JHEP} {\bfseries
  01} (2007) 021} [\href{https://arxiv.org/abs/hep-ph/0611266}{{\ttfamily
  hep-ph/0611266}}].

\bibitem{Steinhauser:1996wy}
M.~Steinhauser, \emph{{Corrections of O $(\alpha_s^{2})$ to the decay of an
  intermediate mass Higgs boson into two photons}},  in \emph{{Ringberg
  Workshop: The Higgs Puzzle - What can We Learn from LEP2, LHC, NLC, and
  FMC?}}, pp.~177--185, 12, 1996,
  \href{https://arxiv.org/abs/hep-ph/9612395}{{\ttfamily hep-ph/9612395}}.

\bibitem{Maierhofer:2012vv}
P.~Maierh\"ofer and P.~Marquard, \emph{{Complete three-loop QCD corrections to
  the decay $H \to \gamma \gamma$}},
  \href{https://doi.org/10.1016/j.physletb.2013.02.040}{\emph{Phys. Lett. B}
  {\bfseries 721} (2013) 131}
  [\href{https://arxiv.org/abs/1212.6233}{{\ttfamily 1212.6233}}].

\bibitem{Niggetiedt:2020sbf}
M.~Niggetiedt, \emph{{Exact quark-mass dependence of the Higgs-photon form
  factor at three loops in QCD}},
  \href{https://doi.org/10.1007/JHEP04(2021)196}{\emph{JHEP} {\bfseries 04}
  (2021) 196} [\href{https://arxiv.org/abs/2009.10556}{{\ttfamily
  2009.10556}}].

\bibitem{Sturm:2014nva}
C.~Sturm, \emph{{Higher order QCD results for the fermionic contributions of
  the Higgs-boson decay into two photons and the decoupling function for the
  $\overline{\text{MS}}$ renormalized fine-structure constant}},
  \href{https://doi.org/10.1140/epjc/s10052-014-2978-0}{\emph{Eur. Phys. J. C}
  {\bfseries 74} (2014) 2978}
  [\href{https://arxiv.org/abs/1404.3433}{{\ttfamily 1404.3433}}].

\bibitem{Davies:2021zbx}
J.~Davies and F.~Herren, \emph{{Higgs boson decay into photons at four loops}},
  \href{https://doi.org/10.1103/PhysRevD.104.053010}{\emph{Phys. Rev. D}
  {\bfseries 104} (2021) 053010}
  [\href{https://arxiv.org/abs/2104.12780}{{\ttfamily 2104.12780}}].

\bibitem{Beneke:1998ui}
M.~Beneke, \emph{{Renormalons}},
  \href{https://doi.org/10.1016/S0370-1573(98)00130-6}{\emph{Phys. Rept.}
  {\bfseries 317} (1999) 1}
  [\href{https://arxiv.org/abs/hep-ph/9807443}{{\ttfamily hep-ph/9807443}}].

\bibitem{Broadhurst:1994se}
D.~J. Broadhurst and A.~G. Grozin, \emph{{Matching QCD and HQET heavy - light
  currents at two loops and beyond}},
  \href{https://doi.org/10.1103/PhysRevD.52.4082}{\emph{Phys. Rev. D}
  {\bfseries 52} (1995) 4082}
  [\href{https://arxiv.org/abs/hep-ph/9410240}{{\ttfamily hep-ph/9410240}}].

\bibitem{Beneke:1994qe}
M.~Beneke and V.~M. Braun, \emph{{Naive nonAbelianization and resummation of
  fermion bubble chains}},
  \href{https://doi.org/10.1016/0370-2693(95)00184-M}{\emph{Phys. Lett. B}
  {\bfseries 348} (1995) 513}
  [\href{https://arxiv.org/abs/hep-ph/9411229}{{\ttfamily hep-ph/9411229}}].

\bibitem{Steinhauser:2000ry}
M.~Steinhauser, \emph{{MATAD: A Program package for the computation of MAssive
  TADpoles}},
  \href{https://doi.org/10.1016/S0010-4655(00)00204-6}{\emph{Comput. Phys.
  Commun.} {\bfseries 134} (2001) 335}
  [\href{https://arxiv.org/abs/hep-ph/0009029}{{\ttfamily hep-ph/0009029}}].

\bibitem{Ruijl:2017dtg}
B.~Ruijl, T.~Ueda and J.~Vermaseren, \emph{{FORM version 4.2}},
  \href{https://arxiv.org/abs/1707.06453}{{\ttfamily 1707.06453}}.

\bibitem{Djouadi:2005gi}
A.~Djouadi, \emph{{The Anatomy of electro-weak symmetry breaking. I: The Higgs
  boson in the standard model}},
  \href{https://doi.org/10.1016/j.physrep.2007.10.004}{\emph{Phys. Rept.}
  {\bfseries 457} (2008) 1}
  [\href{https://arxiv.org/abs/hep-ph/0503172}{{\ttfamily hep-ph/0503172}}].

\bibitem{Smirnov:2002pj}
V.~A. Smirnov, \emph{{Applied asymptotic expansions in momenta and masses}},
  {\emph{Springer Tracts Mod. Phys.} {\bfseries 177} (2002) 1}.

\bibitem{Nogueira:1991ex}
P.~Nogueira, \emph{{Automatic Feynman graph generation}},
  \href{https://doi.org/10.1006/jcph.1993.1074}{\emph{J. Comput. Phys.}
  {\bfseries 105} (1993) 279}.

\bibitem{Beneke:1994sw}
M.~Beneke and V.~M. Braun, \emph{{Heavy quark effective theory beyond
  perturbation theory: Renormalons, the pole mass and the residual mass term}},
  \href{https://doi.org/10.1016/0550-3213(94)90314-X}{\emph{Nucl. Phys. B}
  {\bfseries 426} (1994) 301}
  [\href{https://arxiv.org/abs/hep-ph/9402364}{{\ttfamily hep-ph/9402364}}].

\bibitem{Ball:1995ni}
P.~Ball, M.~Beneke and V.~M. Braun, \emph{{Resummation of (beta0 alpha-s)**n
  corrections in QCD: Techniques and applications to the tau hadronic width and
  the heavy quark pole mass}},
  \href{https://doi.org/10.1016/0550-3213(95)00392-6}{\emph{Nucl. Phys. B}
  {\bfseries 452} (1995) 563}
  [\href{https://arxiv.org/abs/hep-ph/9502300}{{\ttfamily hep-ph/9502300}}].

\bibitem{Beneke:2021lkq}
M.~Beneke, \emph{{Pole mass renormalon and its ramifications}},
  \href{https://doi.org/10.1140/epjs/s11734-021-00268-w}{\emph{Eur. Phys. J.
  ST} {\bfseries 230} (2021) 2565}
  [\href{https://arxiv.org/abs/2108.04861}{{\ttfamily 2108.04861}}].

\bibitem{Grozin:2003gf}
A.~G. Grozin, \emph{{Renormalons: Technical introduction}},
  \href{https://arxiv.org/abs/hep-ph/0311050}{{\ttfamily hep-ph/0311050}}.

\bibitem{Boito:2021wbj}
D.~Boito, V.~Mateu and M.~V. Rodrigues, \emph{{Small-momentum expansion of
  heavy-quark correlators in the large-\ensuremath{\beta}$_{0}$ limit and
  \ensuremath{\alpha}$_{s}$ extractions}},
  \href{https://doi.org/10.1007/JHEP08(2021)027}{\emph{JHEP} {\bfseries 08}
  (2021) 027} [\href{https://arxiv.org/abs/2106.05660}{{\ttfamily
  2106.05660}}].

\bibitem{Boito:2018rwt}
D.~Boito, P.~Masjuan and F.~Oliani, \emph{{Higher-order QCD corrections to
  hadronic $\tau$ decays from Pad\'e approximants}},
  \href{https://doi.org/10.1007/JHEP08(2018)075}{\emph{JHEP} {\bfseries 08}
  (2018) 075} [\href{https://arxiv.org/abs/1807.01567}{{\ttfamily
  1807.01567}}].

\bibitem{Benitez-Rathgeb:2022yqb}
M.~A. Benitez-Rathgeb, D.~Boito, A.~H. Hoang and M.~Jamin, \emph{{Reconciling
  the contour-improved and fixed-order approaches for \ensuremath{\tau}
  hadronic spectral moments. Part I. Renormalon-free gluon condensate scheme}},
  \href{https://doi.org/10.1007/JHEP07(2022)016}{\emph{JHEP} {\bfseries 07}
  (2022) 016} [\href{https://arxiv.org/abs/2202.10957}{{\ttfamily
  2202.10957}}].

\end{thebibliography}\endgroup

\end{document}